\documentclass[sigconf]{acmart}
\acmConference[ESEC/FSE 2023]{The 31st ACM Joint European Software Engineering Conference and Symposium on the Foundations of Software Engineering}{11 - 17 November, 2023}{San Francisco, USA}
\AtBeginDocument{%
  \providecommand\BibTeX{{%
    \normalfont B\kern-0.5em{\scshape i\kern-0.25em b}\kern-0.8em\TeX}}}

\usepackage{microtype}
\usepackage[inline]{enumitem}
\usepackage{xcolor}
\usepackage{xspace}
\usepackage{booktabs} 
 \usepackage{censor}

\newcommand{\eg}{\emph{e.g.,}\xspace}
\newcommand{\ie}{\emph{i.e.,}\xspace}
\newcommand{\etal}{\emph{et al.}\xspace}

\newcommand{\sbsecref}[1]{Subsection~\ref{#1}\xspace}
\newcommand{\figref}[1]{Fig.~\ref{#1}\xspace}

\newcommand{\approach}{\texttt{Idaka}\xspace}

\usepackage[htt]{hyphenat}
\usepackage{todonotes}
\usepackage{menukeys}

\begin{document}

\title{On Using Information Retrieval to Recommend  Machine Learning Good Practices for Software Engineers}



\author{Laura Cabra-Acela}
\affiliation{%
  \institution{Universidad de Los Andes}
  \city{Bogot\'a}
  \country{Colombia}}
\email{lh.cabra@uniandes.edu.co}
\orcid{0009-0000-0121-8560}

\author{Anamaria Mojica-Hanke}
\affiliation{%
  \institution{University of Passau}
  \city{Passau}
  \country{Germany}
}
\affiliation{%
  \institution{Universidad de los Andes}
  \city{Bogot\'a}
  \country{Colombia}
}
\email{ai.mojica10@uniandes.edu.co}
\orcid{0000-0002-5292-2977}

\author{Mario Linares-V\'asquez}
\affiliation{%
  \institution{Universidad de Los Andes}
  \city{Bogot\'a}
  \country{Colombia}}
\email{m.linaresv@uniandes.edu.co}
\orcid{0000-0003-0161-2888}

\author{Steffen Herbold}
\affiliation{%
  \institution{University of Passau}
  \city{Passau}
  \country{Germany}
}
\orcid{0000-0001-9765-2803}
\email{steffen.herbold@uni-passau.de}



\begin{abstract}
  Machine learning (ML) is nowadays widely used for different purposes and in several disciplines. From self-driving cars to automated medical diagnosis, machine learning models extensively support users' daily activities, and software engineering tasks are no exception.  Not embracing good ML practices may lead to pitfalls that hinder the performance of an ML system and potentially  lead to unexpected results. Despite the existence of  documentation and literature about ML best practices,  many non-ML experts turn towards gray literature like blogs and Q\&A systems when looking for help and guidance when implementing ML systems. To better aid users in distilling relevant knowledge from such sources, we propose a recommender system that recommends ML practices based on the user's context. \textit{As a first step in creating a recommender system for machine learning practices, we implemented \approach}. A tool that provides two different approaches for retrieving/generating ML best practices: i) an information retrieval (IR) engine and ii) a large language model. The IR-engine uses BM25 as the algorithm for retrieving the practices, and a large language model, in our case Alpaca. The platform has been designed to allow comparative studies of best practices retrieval tools. \approach is publicly available at \\\textbf{GitHub}: \url{https://bit.ly/idaka}. \textbf{Video}: \url{https://youtu.be/cEb-AhIPxnM}.


\end{abstract}

\ccsdesc[300]{Computing methodologies~Machine learning}
\ccsdesc[300]{Information systems~Information retrieval}
\ccsdesc[500]{Software and its engineering~Software creation and management}

\keywords{Machine learning, Good practices, Information
retrieval, Large language models}

\maketitle

\section{Introduction}

Nowadays, Machine Learning (ML) is used for various applications, including vehicle automation~\cite{tesla_2023, stilgoe2018machine}, medical applications~\cite{rajkomar2019machine,amazon_2019}, adding to the list of daily live ML applications such as chatbots~\cite{adiwardana_2020} and voice assistance~\cite{siri_team_2017}. Software engineering (SE) is no exception, and ML is also used daily by developers  \cite{zhang2003machine,watson2022systematic}. In addition, studies such as~\cite{sculley2015hidden, amershi2019software, Alshangiti_2019} have shown that ML development poses challenges and describe differences between ML development and traditional software development.

Recognizing that ML development presents particular challenges, another group of literature presents guidelines and practices to avoid or deal with them and how to avoid omitting them (\eg \cite{amershi2019software, MichaelLones2021, wujek2016best, zinkevich_2021}). The amount of resources, including white and grey literature on ML, is vast, and when many non-ML experts are looking for guidance, it can be overwhelming not to know where to look for the guidance needed for a specific task. In parallel, we have to deal with the emergence of Chat Bots based on \textit{large language models} like ChatGPT~\cite{OpenAI_2023} as a way that developers obtain information.  To better support users in distilling relevant knowledge from possible sources, a recommender system that suggests practices based on their needs would be ideal. Therefore, our main contributions are the following:
\begin{itemize}
	\item The proposal of \approach, a research prototype that provides two approaches for retrieving ML practices.  i) A first approach, an IR-based search engine for ML practices, ii) an interface for interacting with a generative artificial intelligence, in our case \textit{Alpaca}~\cite{alpaca2023}. \approach is publicly available at \cite{IDAKA_OA} with an MIT licence. 
	\item The \approach platform as a foundation for a searchable and systematic catalog of best practices for machine learning following the ML pipeline proposed by Amershi \etal~\cite{amershi2019software}. The platform is designed to also carry out comparative studies of tools for retrieving best practices.
\end{itemize}

The remainder of this paper is structured as follows. We discuss the context and related work in Section 2. Then, we describe our proposed approach for a search engine for machine learning best practices in Section 3. Finally, we conclude and give an outlook on future work in Section 4.

\section{RELATED WORK}

A considerable amount of white and gray literature addresses ML topics and applications. Some of that literature relates to challenges \cite{Alshangiti_2019}, pitfalls \cite{biderman2020pitfalls}, practices \cite{serban2020adoption, wujek2016best, zinkevich_2021, goole_pair}, or a combination of those \cite{amershi2019software, MichaelLones2021}. This considerable amount of literature shows many possible data sources for retrieving information about ML issues and guidelines. The following briefly discusses the most relevant related work that presents ML guidelines and practices.

Amershi \etal \cite{amershi2019software} is one example of white literature that presents a series of challenges and practices. In particular, they analyzed how Microsoft employees faced significant challenges and their experiences while doing AI. As a result of this analysis, they got a description of a nine-stage ML workflow (\ie \textit{model requirement, data collection, data cleaning, feature engineering, data labeling, model training, model evaluation, model deployment, and model monitoring}) and an overview of some of the best practices for building software that relied on ML. 
Although  they presented a list of practices  and described them, they did not present a way in which, based on a query or a search, the practices could be retrieved.

The most relevant white literature contribution regarding our is the study by Serban \etal~\cite{serban2020adoption}. They  present a study that discussed the adoption of software engineering best practices in ML. They recognized and listed 29  best practices by retrieving academic and gray literature. Then, they validated the practices by surveying  researchers and developers.  The practices were grouped into six categories: \textit{data}, \textit{team}, \textit{training}, \textit{deployment}, \textit{coding}, and \textit{governance}. In addition, they present the results of the survey, which classified the practices into levels of difficulty (\ie basic, medium, and advanced); effects (\ie Agility, Software Quality, Team Effectiveness, Traceability);  and requirements for trustworthy ML (\ie EU Guidelines for Trustworthy AI \cite{commission_2019}). Our work shares the idea of categorizing and listing practices. However, we classified the practices based on the ML pipeline stages mentioned by Amershi \etal~\cite{amershi2019software}. In this way, we focus on the ML component of the system, from collecting the model requirements to model monitoring.  In addition, we propose an additional  step, in which the practices are not only displayed and organized into categories, but also can be retrieved via an information retrieval model, which could be the basis of a future recommendation engine.

A more interactive way of retrieving AI practices is presented by Google in their ``People + AI Guidebook''~\cite{goole_pair}. On this website, Google presented a series of practices grouped by design patterns (\ie ``design patterns highlight key design opportunities for AI products''). They also provide a group of predefined questions (\ie questions already established and associated with a particular number of practices). However, as mentioned, those questions are already predefined, and users cannot search practices with their queries. This restricts the users from knowing and using the predefined vocabulary to search for relevant practices.  Instead, we want to provide the basis of a  model that retrieves practices based on the users' query. To accomplish this, we propose a tool called \approach that provides two options for retrieving the practices, i) an information retrieval (IR) engine, as previous works have done (\eg retrieve traceability links for software engineers~\cite{lucia2007recovering}, retrieve  existing software components for a requirement~\cite{stierna2003applying}, and find requirements based on the proxy of viewpoints model~\cite{lee2004missing}); and ii) an interface to interact with a generative large language model (GLM). 

\section{Proposed Approach}

With the long-term purpose of implementing approaches and tools that allow researchers and practitioners to have a better knowledge of machine learning best practices, we decided first to prototype a search engine (\approach) specialized in this topic. There are general-purpose search engines such as Google or Bing that index sources with different quality, relevance, and soundness, as well as specialized Q\&A websites, such as the StackExchange sites for machine learning and data science, that provide to users with answers various types of questions (\eg how-to, implementation related). There is no doubt that researchers and practitioners use the aforementioned tools. However, their usefulness and effectiveness in ML research has not yet been proven. Therefore, as a first step required to conduct that validation, having a specialized search engine could help us to show whether existing tools are sufficient to cover the needs of users looking for best practices and recommendations when conducting ML experiments or implementing ML systems.

\begin{figure}[h]
	\centering
	\includegraphics[width=0.8\columnwidth]{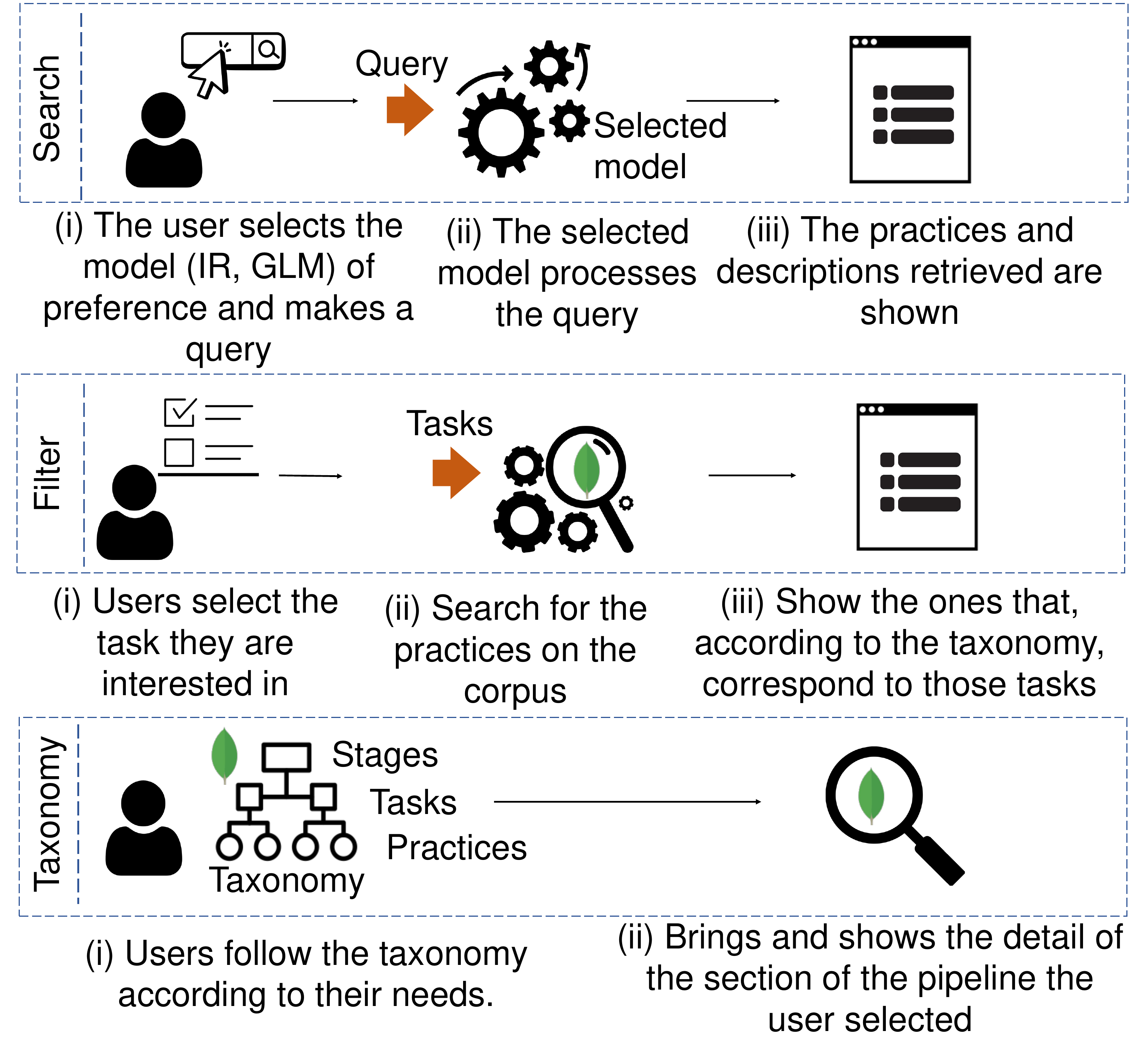}
	\caption{\approach workflow.}
	\label{fig:workflow}
\end{figure}

In general, \approach, besides the option of browsing the practices (see \sbsecref{subsec:Corpus}) by ML stages and ML tasks (\ie browsing in a systematic approach, see \sbsecref{subsec:systematic}), proposes two approaches to answer the users' queries (in search of ML best practices): i) using a classical IR strategy in which documents (ML practices) relevant to a query are retrieved from a corpus based on a similarity metric (see \sbsecref{subsec:IR}); ii) using a generative language model (GLM), like Alpaca~\cite{alpaca2023}, to generate the ML practices (see \sbsecref{subsec:GLM}).  The \approach components and workflow are depicted in \figref{fig:workflow}. 

\vspace{-3pt}

\subsection{Corpus Creation for the IR Engine}
\label{subsec:Corpus}
An IR-based system requires a set of documents that later can be indexed by using a representation such as a bag-of-words~(BoW), vector space model (VSM)~\cite{salton1975vector} and topic modeling~\cite{jiang2018latent}. In the case of our IR, we decided to use sentences in English describing ML best practices as the base to create the required BoW. To create a data set of those sentences, we used an existing data set that contained practices that are extracted by, first selecting relevant Stack Exchange websites. Then,  posts related to ML were extracted from the selected pages. After this, an open-coding process was executed, followed by a practice validation by ML experts~\cite{mojicapp}.  In addition to the previously mentioned practices,  we manually extracted ML best practices (\ie we copied the title, the summary and the description for each practice) from the People + AI Guidebook~\cite{goole_pair}. In total, we obtained a corpus composed of 150 different English sentences describing ML practices.

\subsection{Systematic Approach}
\label{subsec:systematic}
In order to facilitate the process of systematically browsing the practices (searching one by one), 
we grouped the practices into the different ML stages proposed by Amershi \etal~\cite{amershi2019software} (\ie Model requirements, Data Collection, Data Cleaning, Feature Engineering, Data labeling, Model Training, Model Evaluation, Model Deployment, and Model Monitoring), including an additional stage (Support tasks) that relates to practices that support the one or more stages in the pipeline. In addition, we also grouped them in a second level of abstraction (tasks). These are specific activities that can be executed in each stage, \eg handling missing data in the Data Cleaning stage.  \textit{In this way, if a user wants to browse practices by a specific stage or ML task, a more refined search can be done.} 
 
\subsection{Information Retrieval Engine}
\label{subsec:IR}
Before the sentences were indexed using the IR-model, they were pre-processed as follows: i) remove unicode characters; ii) format the words into lowercase; iii) remove punctuation and numbers; iv) remove double spaces, \ie regex $\setminus s\{2,\}$; v) remove stopwords with nltk~\cite{bird2009natural}; and vi) stem the words with Porter stemmer~\cite{porter1980algorithm} provided by nltk~\cite{bird2009natural}. Furthermore, for each sentence, all the nouns and adjectives were changed for their corresponding synonyms as a process of corpus expansion. Afterward, each sentence $s$ in our corpus had a set of terms $V(s)$. The union over all the sets of terms in the sentences describes the vocabulary $V$, which can be used to build different IR models, such as \textit{BM25}~\cite{robertson2009probabilistic}, \textit{Vector Space Model (VSM)}~\cite{salton1975vector}, and \textit{Latent Dirichlet Allocation (LDA)}~\cite{blei2003latent}. In this case, \approach uses \textit{BM25}~\cite{robertson2009probabilistic} as the engine for retrieving practices, based on a users' query. 
\emph{We use \textit{BM25} because, despite being relatively old, it is still widely applied for text retrieval (\eg~\cite{ezzini2023ai, kolthoff2023data, Connelly_2019}) due to its simple implementation and robust behavior~\cite{thakur2021beir}.} 
 
 Note that user queries are also transformed into a BoW, in order to be used by the \textit{IR} model. Therefore, having a query and a corpus indexed with a given model, the most relevant  documents (\ie ML best practices) are retrieved by selecting the top similar ones.
 
For Idaka's IR-Model, we used  \textit{Okapi BM25}, a probabilistic model that computes relevant documents based on the BoW concept. Hence a document's relevance is calculated by considering the frequency of the query terms on it. The following formula is used to calculate the similarity between a document, $d$, and the query, $q$.

$$sim(d,q)= \sum_{i=1}^{n}{IDF(q_i)\cdot \frac{f(q_i,d)\cdot(k_1+1)}{f(q_i,d)+k_1\cdot(1-b+b\cdot\frac{ \| d \| }{avgdl})} } . $$
Given $q_i$ as a term on a query, $f(q_i, d)$ is the number of appearances of $q_i$ on the document, $\| d \|$ is the length of the document, and $avgdl$ represents the average of the documents' length in the collection. Finally, $k_1$ and $b$ are parameters to set depending on the corpus.

\subsection{Generative Language Models}
\label{subsec:GLM}

\textit{Generative Language Models (GLM) are natural language processing models that create new content based on existing ones. They are trained to generate human language according to patterns, structures, and intents in the data supplied}~\cite{cao2023comprehensive}.

For instance, Alpaca~\cite{alpaca2023} is a GLM fine-tuned from Meta's LLaMA 7B model~\cite{touvron2023llama}. Alpaca is trained to generate outputs based on instructions given by the users, behaving qualitatively similar to other existing services, such as  Bloom~\cite{workshop2023bloom}, Chinchilla~\cite{hoffmann2022training}, BERT~\cite{BERT}, and ChatGPT service, which serves as Web front-end to GPT-3.5~\cite{ouyang2022training} and GPT-4~\cite{openai2023gpt4}.

Following this idea and based on the user's query, \approach utilizes Alpaca~\cite{alpaca2023} (\textit{publicly available}) to recommend good practices for machine learning tasks. It uses the NodeJS API Dalai~\cite{Dalai} to run Alpaca locally and send the corresponding requests.  Since Alpaca was not trained to reply with a list of practices but to generate new sentences based on the information that is being used as input (\ie \texttt{users' query}),  we needed to use prompt engineering to force Alpaca to reply i) practice(s), and not only a completion of the \texttt{users' query}; and ii) a description of the practices. In this way, we tried to ensure a similar response to the structure already established by the IR corpus (\ie practice + description). We used an engineered prompt~\cite{radford2019language, liu2023pre} which we prepended to all the queries to achieve this. As follows:

\textbf{\texttt{Prompt:}} \texttt{Below is an instruction that describes a task, paired with an input that provides further context. Write a response that appropriately completes the request.\#\#\# Instruction: Give me an enumerated list of best practices for + \textbf{user's query} $+$ with a description of each of them.}

\subsection{Tool Implementation}

We implemented \approach as a web system publicly available online~\cite{idaka_render, IDAKA_OA}, including a video demonstration~\cite{idaka_tool_demo}. The technology stack that we used is the following: \textit{React.js} (front-end), \textit{Express.js} (back-end), \textit{Node.js} for the application back-end including the Dalai-API for the GLM, \textit{Python} for the IR model, and \textit{Atlas MongoDB} for storing the IR corpus and user feedback.

\begin{figure}
	\centering
	\includegraphics[width=0.88\columnwidth]{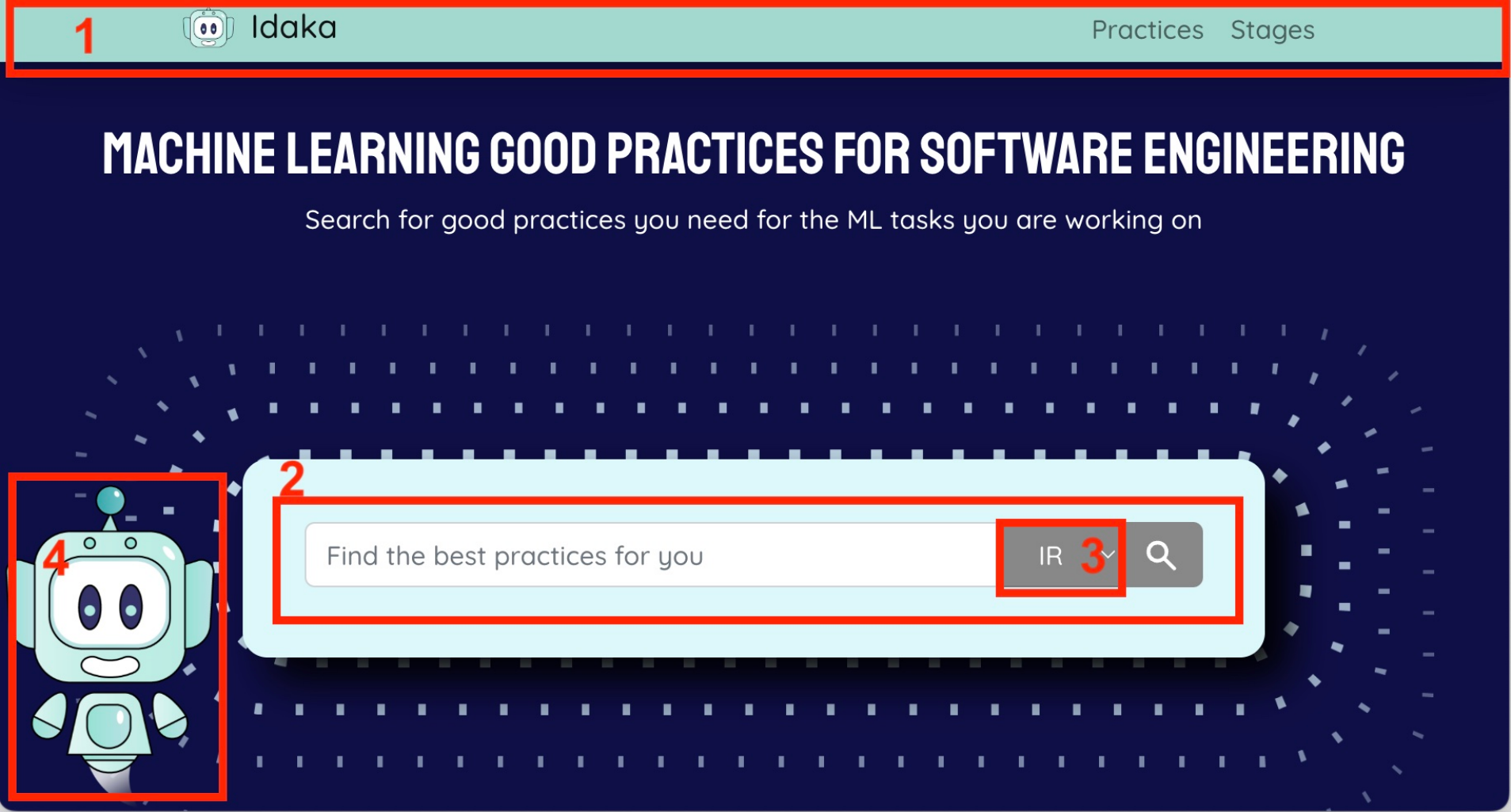}
 	\caption{\approach tool: home view}
	\label{fig:main_view}
\vspace{-10pt}
\end{figure}

\begin{figure}
	\centering
	\includegraphics[width=0.88\columnwidth]{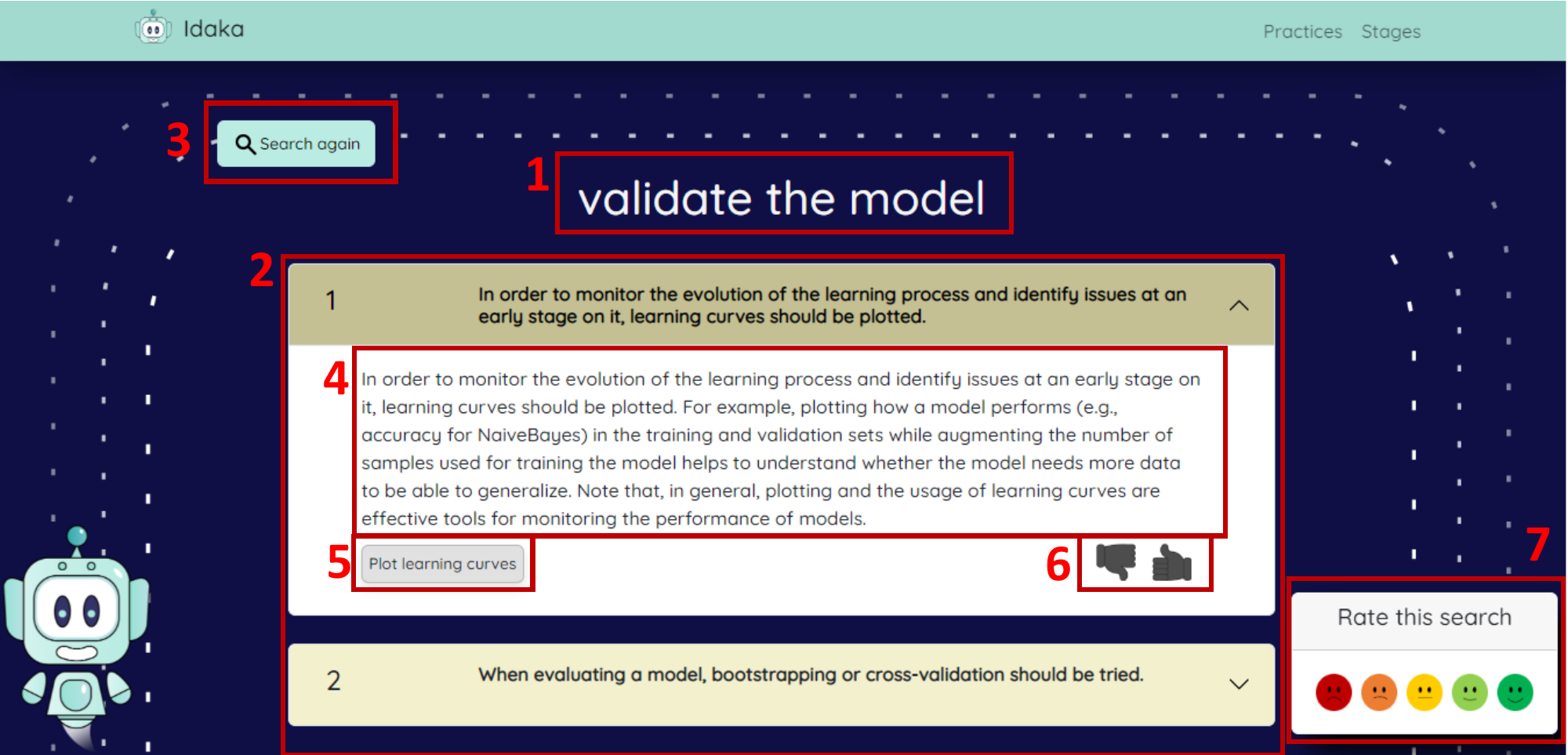}
	\caption{\approach tool:  results view}
	\label{fig:results_view}
\vspace{-10pt}
\end{figure}

\begin{figure}
	\centering
	\includegraphics[width=0.88\columnwidth]{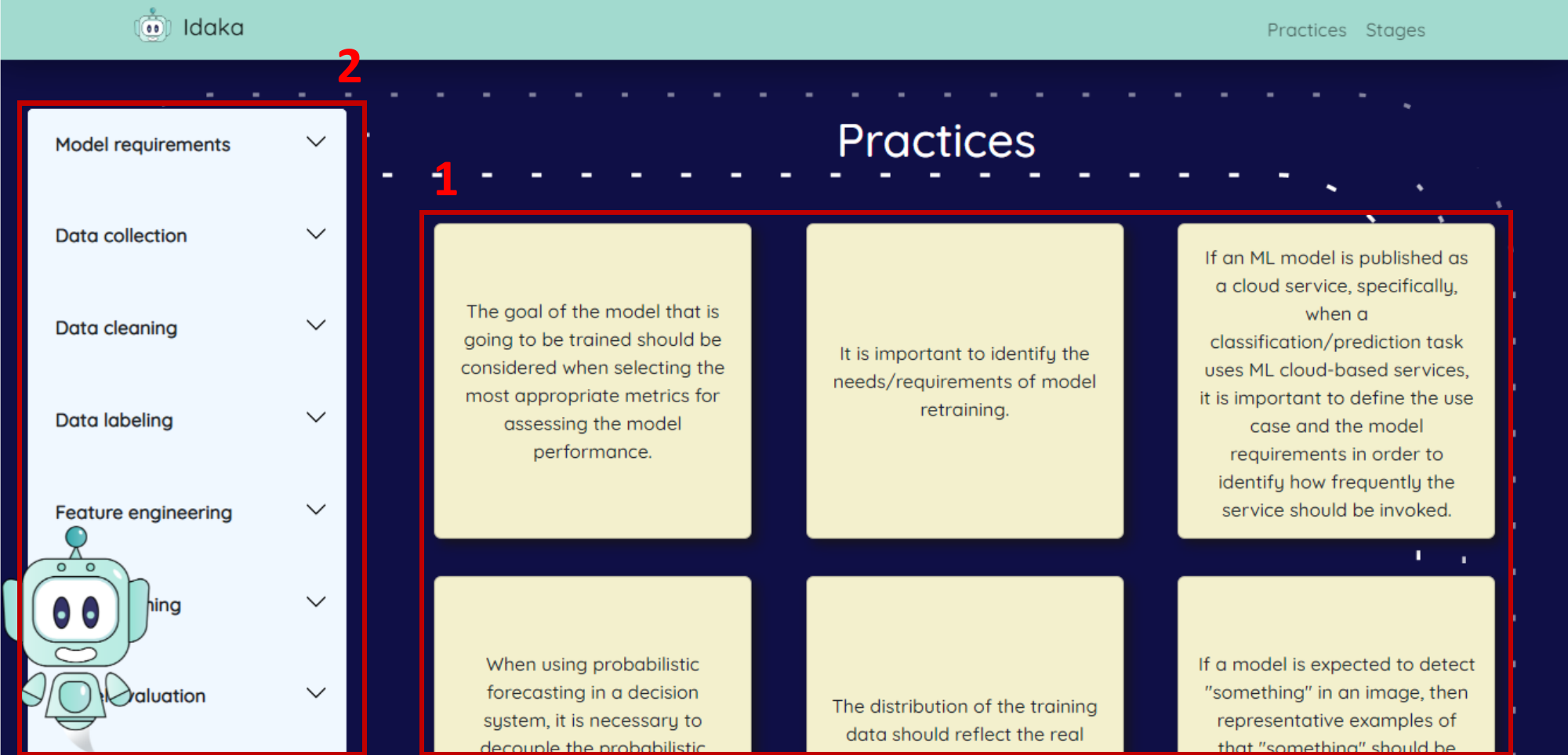}
	\caption{\approach tool: practices view}
	\label{fig:practices_view}
\vspace{-10pt}
\end{figure}

\begin{figure}
	\centering
	\includegraphics[width=0.88\columnwidth]{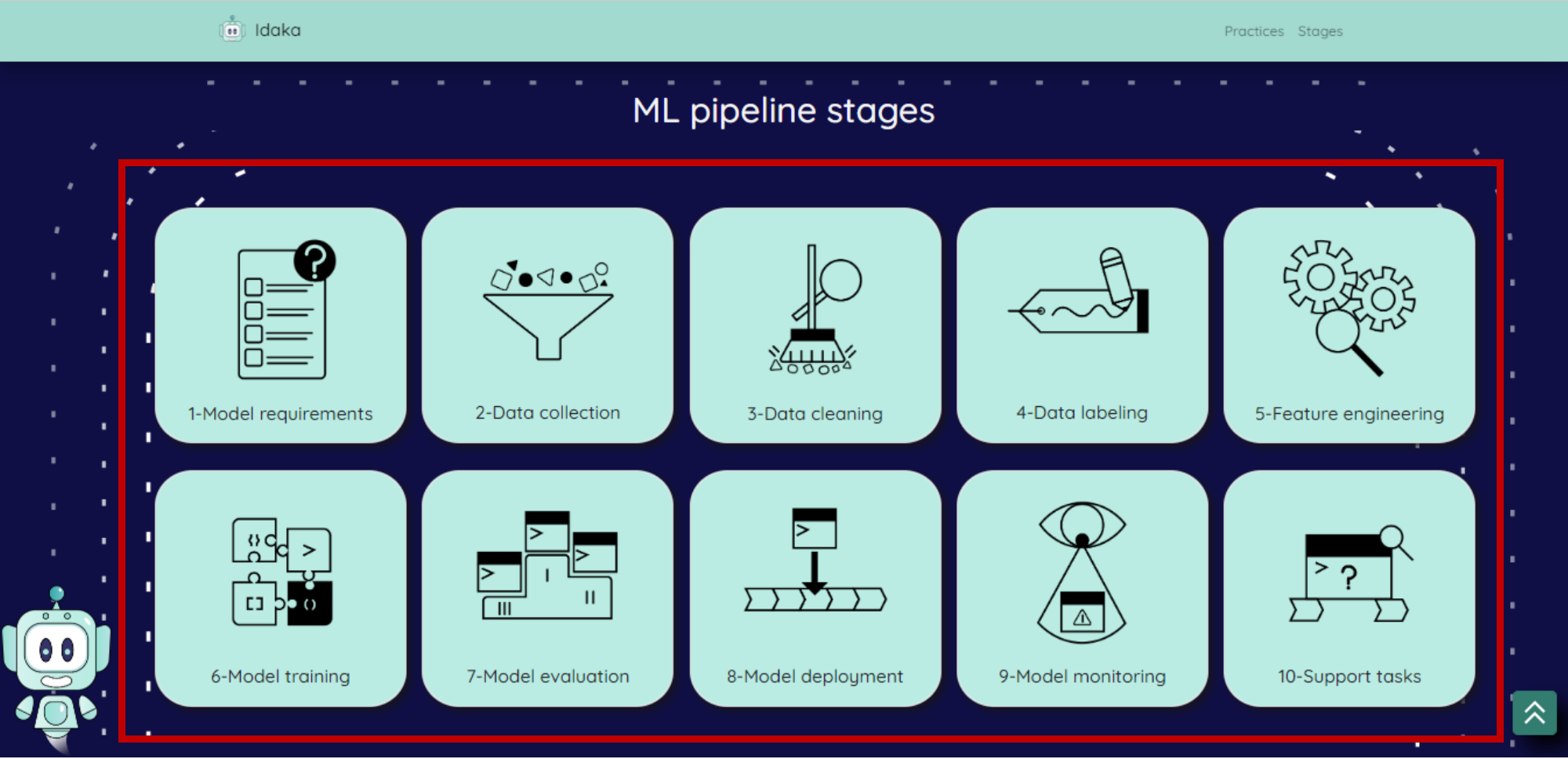}  	
	\caption{\approach tool:  stages view}
	\label{fig:stages_view}
 \vspace{-15pt}
\end{figure}

\approach provides users with four main views: \textit{home} (\figref{fig:main_view}), \textit{results} (\figref{fig:results_view}), \textit{practices} (\figref{fig:practices_view}), and \textit{stages} (\figref{fig:stages_view}), which can be accessed from the home view with a navigation bar (\figref{fig:main_view} (1)). In the home view (\figref{fig:main_view}), users can type a query in the search bar (2). When querying the practices, the user can select between the two options provided by \approach (\ie by an information retrieval engine or using a generative language model). In addition, there is the \approach robot avatar (4)  which guides the user on the whole web page; once a user gets to each view, a dialog from the robot appears, telling the user what she can do on a  view.

The results view (\figref{fig:results_view}) shows the practices relevant (2) to the query  (1), sorted by similarity (in the case of the IR model). Each listed practice has an augmented description (4), an ML task the practice belongs to (5), and two buttons to report whether the retrieved practice is useful. Note that (3) and (4) are not always available. In particular, for (3)  \textit{Alpaca} does not always gives a description of the practices, and since the practices are created on the fly, those are not associated to a particular task (4).

The practices view  (\figref{fig:practices_view}) lists the whole catalog of practices in the corpus.
The user can select filters (2) to get a refined list of practices (1). 
Finally, the stages view (\figref{fig:stages_view}) groups the practices based on the ML pipeline 
 (Amershi \etal \cite{amershi2019software}).  The list of practices and their stages is only available for the IR practice corpus.

\section{Conclusion \& Future Work}

We expect \approach in the future will facilitate access to good ML practices for SE, helping to avoid pitfalls and challenges that hinder the achievement of good performance of ML-enabled systems. Since the proposed approach is in its very first version, there are several opportunities for improvement and evaluation. 

As part of our work, we are considering how we can be more proactive in supporting developers and researchers that use ML. The available catalogs of practices have the limitation that they require a great up-front effort by the developers to read through the complete catalogs to be aware of them, \eg \cite{amershi2019software, serban2020adoption, MichaelLones2021}. While this is certainly the ideal scenario, they often rather look for solutions to specific problems, as is highlighted by the popularity of platforms like StackOverflow. We enable this, \ie searching  ML practices based on a need, through an IR-based approach to recommend best practices for ML.  \approach enables the browsing of a curated catalog of best practices, and the retrieval of practices based on users' queries. 

In addition, for researchers, besides having a tool for searching ML practices, we implemented this tool in a way that could be used to carry out comparative studies of tools for retrieving best practices by implementing a tool that allows i) effortless change between the back-end (\ie IR model or Alpaca) being used for answering the users' query; and ii) by homogenizing the output of both models, in this way for an end-user both approaches answer in a similar way (\ie practice + description) with the same structure provided by \approach, helping to mitigate biases in tools-comparative studies.

For future work, we need to enhance the IR approach to be able to deal with intrinsic characteristics of the written language, such as polysemy, homonyms,  and typos. Regarding the GLM, we should work on its performance, which will imply the usage of a bigger infrastructure to improve response times. This should also be improved for using this tool for conducting tool-comparative studies. In addition, options that facilitate executing these kinds of studies should be enabled, like hiding the back-end used for answering the users’ queries. Finally, while we have already added a comprehensive corpus of 150 practices to the IR \approach, we plan to systematically extend this corpus based on additional sources from the literature, \eg \cite{amershi2019software, serban2020adoption}.

\section{Tool Availability}
\approach is publicly available as a website~\cite{IDAKA_OA}, and in a long term archive~\cite{IDAKA_ZENODO}. Inside the repository, it is possible to find code, instructions, and data to build and deploy the tool. The data, as previously mentioned, was the result of collection ML practices from existing data sets~\cite{mojicapp, goole_pair}. 


\bibliographystyle{ACM-Reference-Format}
\bibliography{1_local_bib}

\end{document}